\documentclass[preprint]{aastex}
\usepackage{graphicx}
\begin{document}

\title{Nonlinear propagation of planet-generated tidal waves.}
\author{R. R. Rafikov}
\affil{Princeton University Observatory, Princeton, NJ 08544}
\email{rrr@astro.princeton.edu}

\begin{abstract}
The propagation and evolution of planet-generated density waves
in protoplanetary disks is considered.
The evolution of waves, leading to the shock formation and 
wake dissipation, is followed in the weakly nonlinear regime.
The local approach of Goodman \& Rafikov (2001) is extended to include 
 the effects of surface density and 
temperature variations in the disk
 as well as  the disk 
cylindrical geometry and nonuniform shear.
Wave damping due to shocks 
is demonstrated to be a nonlocal process spanning a significant
fraction of the disk.
Torques induced by the planet could be significant drivers of disk 
evolution  on timescales $\sim 10^6-10^7$ yr even in the absence of strong 
background viscosity.  
A global prescription for 
angular momentum deposition is developed which could be incorporated into 
 the study of gap formation in a gaseous disk around the planet.
\end{abstract}


\section{Introduction.}\label{intro}

Tidal disk-companion interactions are important in 
a variety of astrophysical contexts ranging from the 
formation and evolution of 
protoplanetary systems to the origin of galactic spiral structure. 
Gravitational
interaction between the disk and companion generates density waves
in the disk (gaseous, stellar or particulate) 
which carry angular momentum.
This angular momentum is eventually deposited elsewhere in the disk,
 leading to the evolution of 
the disk as a whole.

In the case of protoplanetary systems disk-planet interactions
not only cause the migration of the planet (Goldreich \& Tremaine 1980, 
hereafter GT80; Ward 1997) but can also drive 
noticeable evolution of
the disk itself (Larson 1989; Goodman \& Rafikov 2001, 
hereafter GR01). 
However, to change the 
state of the disk and orbit of the planet in these systems 
it is necessary to somehow transfer angular momentum
from the density waves launched by the perturber 
to the disk material and this can only be accomplished 
by virtue of some damping process (Goldreich \& Nicholson 1989). 

Various mechanisms have been envisaged as possible sources for such  
damping. The most popular linear ones are viscosity in the disk 
(Takeuchi et al. 1996)
and radiative
damping of the tidal perturbations (Cassen \& Woolum 1996). 
Viscosity can dissipate tidal perturbations on scales smaller than the 
typical disk sizes only if it is
large enough, $\alpha\ga 10^{-4}$ (Takeuchi et al. 1996; GR01), 
but it is hard to identify a strong source of 
viscosity in protoplanetary disks. The most probable viscous 
mechanism in hot accretion disks -- magnetohydrodynamic (MHD)
turbulence driven by the magnetorotational instability (Velikhov 1959; 
Balbus \& Hawley 1998) -- probably does not operate
 in protoplanetary disk
environments: the gas is too cold and weakly ionized
throughout most of the disk (Jin 1996; Hawley \& Stone 1998). 
Convection was put forward as another possible source of
viscosity (Lin \& Papaloizou 1980) but analytical arguments and 
numerical simulations cast serious doubt on the ability of this mechanism 
to  produce outward angular momentum transport 
 (Ryu \& Goodman 1992; 
Balbus et al. 1996).
The efficiency of radiative damping in protoplanetary disks
 is strongly reduced by 
 dust opacity  (Henning \& Stognienko 1996), which
leads to 
very high optical depths ($\tau\ga 10^3$) and 
implies very small radiative losses from the disk surface. Waves could also 
be damped by radiative transfer in the plane of the disk 
 but this turns out not to be very important either (GR01).

Nonlinear dissipation, namely shock formation and its consequent 
damping, 
seems to represent a more efficient and almost inevitable 
process for transferring the angular 
momentum of the wave to the disk fluid.
There are two reasons for this. First, the differential rotation of
the background fluid causes the wavelength of the tidal perturbation to
decrease as it travels away from the planet
(Goldreich \& Tremaine 1979). Second, the amplitude 
of the planet-induced wake is growing, at least in the planetary vicinity,
as a consequence of the conservation of the angular momentum flux.
These processes working together can make the wave shock very rapidly 
if the initial wave amplitude is significant (i.e. if the perturber is
massive enough).

Goodman \& Rafikov (GR01) 
have performed a detailed study of   nonlinear evolution
of planet-induced density waves in two-dimensional disks. Using 
the shearing sheet approximation 
they have demonstrated that for small enough planetary mass
 it is possible to
separate the wake evolution into two distinct stages: linear generation, which
takes about $(1-2)\times h$ from the planet to complete (here $h=c/\Omega$
is a typical scale length, which equals  disk 
thickness in three dimensions, 
$c$ is a sound speed and $\Omega$ is a disk angular frequency), and
then nonlinear evolution of the 
wake causing it to shock after travelling  
several $h$ from the perturber (of course, depending
on the mass of the planet). After the shock is formed, it damps
transferring its angular momentum to the mean flow and leading to
disk evolution. 
In favorable conditions
the damping of the density waves can make the
disk evolve on timescales comparable with those derived from observations
[$10^{5}-10^{7}$ yr, see Hartmann et al. (1998)].

The considerations of GR01 were in a 
certain sense local because of the shearing sheet approximation. 
It was shown that this approach
 is good for the description of the shock {\it formation}
but probably not so accurate for studying the shock {\it dissipation}: 
in the shearing
sheet geometry shock damping proceeds slowly and at some point the
background fluid velocity can no longer be represented by just a uniform 
shear. 
The analysis of GR01 also assumed  a disk
with constant background surface density and sound speed and 
did not take into account  the effects
of the disk's polar geometry on the evolution of the density wave amplitude. 
These approximations naturally lead
to a picture in which the wake itself and its damping pattern are 
symmetric on both sides of the planet. 

The main purpose of this paper is to
 extend the analysis of GR01 by including the effects of
radial surface density and sound speed variations in the disk
on the
behavior of the weak tidal disturbances generated by a low-mass planet
incapable of opening a gap in the disk. We consider a Keplerian rotation 
law (not a linearly sheared background flow) 
and a polar geometry to include self-consistently
important ingredients needed to provide a global picture of the 
nonlinear evolution of the density waves in 
non self-gravitating disks. We restrict our 
attention to purely two-dimensional disks, thus completely disregarding
vertical motions and related phenomena, such as wave-action
channeling in the vertical direction 
(Lin et al. 1990; Lubow \& Ogilvie 1998). We believe that this is a 
good approximation in passive, externally irradiated protoplanetary disks
with high optical depths which should have almost isothermal vertical 
structure (Chiang \& Goldreich 1997). 

The paper is structured as follows.
In \S \ref{eqns} we study the full system of fluid equations in 
cylindrical geometry. We provide 
in \S \ref{gencons}
simple scaling laws for the behavior of the wake amplitude and 
shock formation distance in the 
quazilinear approximation from rather general qualitative considerations. 
We then confirm these estimates by an accurate quantitative analysis 
in \S \ref{basder}.
Wake properties are studied in \S \ref{wakeprop}, in particular for the case 
of disks with power-law surface density and sound speed radial dependencies
(\S \ref{pow-law-disks}). 
Applications of our results are discussed in \S \ref{disc}.

\section{Density wave structure.}\label{eqns}

We consider a system consisting of a 
gaseous non self-gravitating
disk rotating in the unperturbed potential $U_\star(r)$
($r$ is the distance from the central object)
and a planet of mass $M_p$ located
at a distance $r_p$ from the center. The disk is assumed to be 
geometrically thin and its unperturbed surface density $\Sigma_0$
and sound speed $c_0$
are both taken to be functions of $r$.
Disk scale height in three dimensions 
is $h=c_0/\Omega$. We will always denote by 
the subscript ``p''
various quantities evaluated at the position of the planet.  

The equations of motion and continuity for the 
two-dimensional disk read as always
\begin{eqnarray}
\frac{\partial {\bf v}}{\partial t}+{\bf v}\cdot\nabla{\bf v}=
-\frac{1}{\Sigma}\nabla P-\nabla U,\label{moteq}\\
\frac{\partial \Sigma}{\partial t}+\nabla\cdot
\left({\bf v}\Sigma\right)=0.
\label{conteq}
\end{eqnarray}
Here ${\bf v}$ and $\Sigma$ are the  fluid velocity
and surface density, $P$ is the pressure, and the potential 
\begin{equation}
U=U_\star-G\frac{M_p}{|{\bf r}-{\bf r}_p|}+U_i,\label{pot}
\end{equation}
consists of three contributions: the potentials of the central star 
and perturbing planet (we assume a Keplerian potential
$U_\star=-GM_c/r$, where $M_c$ is a mass of the central star), 
and the indirect potential $U_i$, which can always be 
neglected here (because disk mass is much smaller than $M_c$).

We assume that  
the pressure $P$ is related  
to the instantaneous value
of the surface density $\Sigma$ by the
{\it locally} polytropic law with a specific index $\gamma$:
\begin{equation}
P=P_0(r)\left[\frac{\Sigma}{\Sigma_0(r)}\right]^\gamma.\label{pol-law}
\end{equation}
Then the perturbed sound velocity is given by the usual expression
\begin{equation}
c^2=\frac{\partial P}{\partial \Sigma}=c_0^2(r)
\left[\frac{\Sigma}{\Sigma_0(r)}\right]^{\gamma-1},
\label{sound_vel}
\end{equation}
meaning that $P_0(r)=\Sigma_0(r) c_0^2(r)/\gamma$. Clearly, $P_0(r)$ and 
$c_0(r)$ are unperturbed values of pressure and sound speed.
The equation of state given by (\ref{pol-law}) does not force $c_0(r)$ to
be related to $\Sigma_0(r)$ and this gives us additional flexibility 
in applications. The entropy
of the disk fluid now varies with $r$, contrary to the equation
of state with a radially-independent polytropic constant. 

\subsection{General considerations.}\label{gencons}

Using basic physical principles such as conservation of 
angular momentum flux it is possible to analyze 
the behavior of the wake in the quazilinear regime and determine when the wave
shocks on a qualitative level.
The simple scaling laws obtained in this way 
will later be confirmed by more rigorous 
analysis of the complete system of the fluid equations.

Let us first concentrate on linear wake propagation. 
One can easily show that the solution of equations 
(\ref{moteq})-(\ref{conteq}) in the linearized form with the
equation of state (\ref{pol-law}) yields, using 
the WKB approximation, 
\begin{equation}
m^2\left(\Omega-\Omega_p\right)^2=\kappa^2+k^2 c^2,
\label{disp}
\end{equation} 
and
\begin{equation}
\frac{d}{dr}\left[\frac{r c_0^3}{(\Omega-\Omega_p)\Sigma_0}
(\Sigma-\Sigma_0)_m^2\right]=0,\label{cons}
\end{equation}
for non self-gravitating disks. Here $m$ is the azimuthal wavenumber, 
$k$ is the radial wavenumber of the 
perturbation, $\Omega_p$ is the perturbation pattern 
angular frequency, $(\Sigma-\Sigma_0)_m$ is the 
$m$-th harmonic of the surface density perturbation, and
\begin{equation}
\Omega^2=\frac{1}{r}\frac{d U_\star}{dr}+\frac{1}{\Sigma_0}
\frac{d P_0}{dr}, ~~~\kappa^2(r)=4B(r)\Omega(r),\label{omegas}
\end{equation}
where
\begin{equation}
B(r)=\Omega(r)+\frac{r}{2}\frac{d\Omega}{dr}.\label{B}
\end{equation}
 
Equation (\ref{disp}) is the usual dispersion relation for small perturbations
in the thin differentially-rotating disk and it shows that density waves behave
basically 
like sound waves after propagating several scale lengths $h$ from the 
perturber.
Both equations (\ref{disp}) and (\ref{cons})
 coincide with their analogs for the 
disks with constant entropy (Goldreich \& Tremaine 1979). One can also
demonstrate that the total angular momentum flux across a cylinder
of radius $r$ carried 
by the $m$-th harmonic of surface density perturbation is given by 
\begin{equation}
f_{J m}=\pi\frac{r c_0^3}{(\Omega-\Omega_p)\Sigma_0}(\Sigma-\Sigma_0)_m^2
\label{fjm-sig},
\end{equation}
[which is the same as the analogous expression of Goldreich \& Tremaine 
(1979)], which, combined with equation(\ref{cons}), implies that $f_J$ 
is conserved.

If the nonlinear effects are fully neglected and 
$f_J$ is strictly conserved as demonstrated 
above, then it follows from (\ref{fjm-sig}) that
the magnitude of the surface density perturbations
scales with other flow variables as
\begin{eqnarray}
(\Sigma-\Sigma_0)^2 \propto\frac{\Sigma_0(\Omega-\Omega_p)}{r c_0^3}f_J.
\label{scaling}
\end{eqnarray}
This equation demonstrates how the amplitude of the wave varies in the linear
regime and we will return to it in \S \ref{basder}.

In reality, even if the perturbation is small,
different points of the wake profile have different propagation 
velocities so that the waveform 
constantly distorts. This nonlinear evolution leads to shock 
formation. In a shock the angular momentum of the density wave
gets transferred to the mean flow so that scaling provided by 
(\ref{scaling}) breaks down after the shock is formed.
We can estimate when this happens. 

Let us consider a part of the profile which initially has a phase 
$\varphi=\varphi_0$. It evolves according to
\begin{equation}
\frac{\partial \varphi}{\partial r} \bigg|_{\varphi_0}=
2\pi \frac{\delta c(r)}{\lambda(r)c(r)},
\label{ev}
\end{equation}
where $\delta c(r)$ is the perturbation of the propagation 
velocity $c$, which is the sound velocity 
in our case (it is different for different
$\varphi_0$ and this is responsible for the profile distortion), and 
$\lambda(r)$ is the current wavelength.

Integrating (\ref{ev}) with respect to $r$ one gets that
\begin{equation}
\varphi=\varphi_0+2\pi\int\limits^r_{r_0}\frac{\delta c(r)}
{\lambda(r)c(r)}dr
\label{varph}
\end{equation}
($r_0$ corresponds to the point where the wave is launched).

The wave shocks when its profile acquires an infinite slope after travelling
some distance $r_{sh}$: 
$d\varphi/d\varphi_0=0$  at $r=r_{sh}$ for some 
specific $\varphi_0$. This gives us [using  
(\ref{varph})] the condition for $r_{sh}$ in the following form:
\begin{equation}
\int\limits^{r_{sh}}_{r_0}\frac{k(r)\delta c(r)}{c(r)}dr=const.
\end{equation}
From the dispersion relation (\ref{disp}) we get that 
$k\propto (\Omega-\Omega_p)/c$; 
since $\delta c/c \propto \delta \Sigma/\Sigma$, we can use the 
scaling provided 
by equation (\ref{scaling}) to obtain finally the shocking condition in the form
\begin{equation}
\int\limits^{r_{sh}}_{r_0}
\left[\frac{(\Omega-\Omega_p)^{3}}{r c_0^{5}\Sigma_0}\right]^{1/2}dr=C 
f_J^{-1/2}\propto M_p^{-1},
\label{shck}
\end{equation}
where $C$ is some constant.

Scaling laws (\ref{scaling}) and (\ref{shck}) immediately provide 
important information about the wake propagation in the linear regime and 
conditions needed for the shock formation.

\subsection{Basic equations.}\label{basder}

In this section 
we study the behavior of the wake by solving the fluid 
equations for small perturbations in a weakly nonlinear regime.
We work in the polar coordinate system rotating
with the angular velocity of the perturber 
$\Omega_p=\sqrt{(1/r_p)(\partial U_\star/\partial r)|_{r_p}}$. 
In this coordinate frame the flow is stationary (time-independent). 
As always, we take the $r$-axis to be directed out from the 
central body and the $\phi$-axis
to be directed in a prograde sense. 
 
In this coordinate system the equations
of motion in the $r$ and $\phi$ directions are (Landau \& Lifshitz 1959)
\begin{eqnarray}
v_r\partial_r v_r+\frac{v_\phi}{r}\partial_\phi v_r-\frac{v_\phi^2}{r}=
-\frac{1}{\Sigma}\partial_r P +2\Omega_p v_\phi+\Omega^2_p r
 -\partial_r U,
\label{genmotr}\\
v_r\partial_r v_\phi+\frac{v_\phi}{r}\partial_\phi v_\phi+\frac{v_\phi v_r}{r}=
-\frac{1}{\Sigma r}\partial_\phi P -
2\Omega_p v_r - \frac{1}{r}\partial_\phi U,
\label{genmotphi}
\end{eqnarray}
and the continuity equation is
\begin{eqnarray}
\frac{1}{r}\partial_r\left(\Sigma r v_r \right)+\frac{1}{r}\partial_\phi
\left(\Sigma v_\phi \right)=0.\label{cont}
\end{eqnarray}
Here $v_r$ and $v_\phi$ are the fluid velocities in the $r$ and $\phi$
directions respectively.
Equations (\ref{genmotr}) and  (\ref{genmotphi}) include Coriolis  and
centrifugal forces.

GR01 have 
demonstrated that the system (\ref{genmotr})-(\ref{cont}) 
simplifies significantly 
if the mass of the perturber 
$M_p$ is smaller than a characteristic mass $M_1$ given by
\begin{equation}
M_1\equiv \frac{c_0^3}{|2A|G}\label{M1},
\end{equation}
where $A(r)=(r/2)d\Omega/dr$ is the Oort's $A$ constant which is related
to $B(r)$ by $B\equiv\Omega + A$ [equation(\ref{B})].
Then the
disk could be split into two distinct regions:
an excitation region, within several scale lengths 
$h_p=h(r_p)=c_0(r_p)/\Omega(r_p)$
from the planet, where one can neglect nonlinear effects, and a
propagation region beyond several $h_p$ from the planet, where 
the planetary potential is negligible and one can  
study wake evolution caused by nonlinear effects. 

In the first region numerous Lindblad resonances tidally excite 
density waves corresponding to different azimuthal 
 harmonics and provide individual Fourier contributions to
the angular momentum flux (GT80). Most of the angular momentum comes from 
the resonances with azimuthal wavenumbers $m\sim r_p/h_p\gg 1$ 
which lie close
to the planet --- at distances $\sim h_p$ from it. Harmonics with smaller
$m$ are weaker because the 
tidal excitation they experience is reduced by their  
larger distance from the planet. Fourier components with 
$m$ higher than $r_p/h_p$ are strongly damped because of the so-called ``torque
cutoff'' (GT80). Its nature can be qualitatively understood as follows: 
near the planet, less than $(2/3)h_p$ from it (in a Keplerian disk), 
the background fluid 
flow is subsonic which does not allow a stationary perturber to excite 
sound waves (Landau \& Lifshitz 1959). This cutoff 
strongly reduces the amount of
the angular momentum flux carried by the corresponding harmonics.

GR01 studied the linear wake formation process in the 
shearing sheet approximation taking into account not only the contributions of
individual resonaces to the torque, but also the phases of the Fourier
harmonics of the surface density perturbation 
which allowed them to obtain the {\it shape} of the wake
in the linear theory.  
Here we will simply use their results in our more general case  
because wake generation is essentially 
a local process, spanning only a few $h_p$ from the planet, where  
the shearing sheet approximation gives a good representation of the 
disk velocity profile and $\Sigma_0$ and $c_0$ may be assumed constant.
The solution of this problem in our more global setting
with varying $\Sigma_0$ and $c_0$ will only be different
from the previously obtained 
 local solution by factors of the order of $h_p/r_p\ll 1$
which is of no interest for us here.

Thus, we can proceed immediately to study the wake propagation region.
To do this we use a simple extension of conventional perturbation theory 
capable of including a weak nonlinearity of the wave. We 
assume that $v_r$ and $v_\phi$ are given by 
\begin{eqnarray}
v_r=u,~~~v_\phi=v_0(r)+v~~~
\mbox{with}~~~ v_0(r)=r\left[\Omega(r)-
\Omega_p\right], \label{as2}
\end{eqnarray}
where $u$ and $v$ are the velocity perturbations and we 
take $u,v\ll v_0$ since the
shock is assumed to be weak and we are always several scale lengths away from 
the planet. We will often write for brevity 
$\Delta \Omega=\Omega(r)-\Omega_p$.

Substituting  (\ref{as2}) into (\ref{genmotr})-(\ref{cont})
one obtains the following perturbation equations: 
\begin{eqnarray}
\Delta \Omega \partial_\phi u+u\partial_r u -2\Omega v
=-\left(\frac{1}{\Sigma}
 \partial_r P-\frac{1}{\Sigma}
 \partial_r P_0\right) +\frac{v^2}{r},\label{linr}\\
\Delta \Omega\partial_\phi v+u\partial_r v +2 B u=-
\frac{1}{r\Sigma}
\partial_\phi P-
\frac{v u}{r},\label{linphi}\\
\Delta \Omega\partial_\phi \Sigma+u\partial_r \Sigma +
\Sigma\partial_r u=
-\frac{\Sigma u}{r}-\frac{v}{r}\partial_\phi \Sigma
-\frac{\Sigma}{r} \partial_\phi v.\label{lincont}
\end{eqnarray}
We have everywhere neglected $v$ in comparison with $\Delta \Omega r$
and made use of the fact that $\partial_\phi P_0=0$. In equations 
(\ref{linr})-(\ref{lincont}) terms quadratic in $u$ and $v$ are subdominant;
however we will keep $u\partial_r u$ terms because they are the strongest 
drivers of nonlinear evolution. 
Also, we assume that $v\ll u$ and $\partial_\phi\ll r\partial_r$
as a consequence of tight-winding approximation. These assumptions are checked
in Appendix.

By introducing a new radial coordinate given by
\begin{equation}
\xi=\int\limits_{r_p}^r \left[\Omega(r)-\Omega_p\right] dr\label{xi}
\end{equation}
this system is transformed into
\begin{eqnarray}
\partial_\phi u+u\partial_\xi u+\frac{1}{\Sigma}\partial_\xi P
-\frac{1}{\Sigma_0}\partial_\xi P_0
=\frac{2\Omega}{\Delta \Omega} v+\frac{v^2}{\Delta \Omega r},\label{linr1}\\
\partial_\phi
 v+u\partial_\xi v +
\frac{c^2}{\Delta \Omega r\Sigma}\partial_\phi \Sigma
=-\frac{2 B}{\Delta \Omega} u-\frac{u v}{\Delta \Omega r},\label{linphi1}\\
\partial_\phi \Sigma+u\partial_\xi \Sigma +
\Sigma \partial_\xi u=
-\frac{\Sigma u}{\Delta \Omega r}-\frac{v}{\Delta \Omega r}\partial_\phi
\Sigma-\frac{\Sigma}{\Delta \Omega r}\partial_\phi v.\label{lincont1}
\end{eqnarray}

The l.h.s. of (\ref{linr1}) and (\ref{lincont1}) are similar to the usual 
system of equations describing isentropic one-dimensional 
gas motion (Landau \& Lifshitz 1959), which possesses two invariants
conserved on characteristics --- Riemann invariants $R_\pm$. 
In our case the nonzero r.h.s. of (\ref{linr1}) and (\ref{lincont1}) 
cause these invariants not to be conserved exactly, but one can 
still use them
in a slightly modified way.
We extract from (\ref{linr1}) and (\ref{lincont1})  two 
equations of evolution of the  Riemann invariants:
\begin{eqnarray}
\left[\partial_\phi+(u\pm c)\partial_\xi\right]R_\pm=
-\left[\frac{1}{\Sigma}\partial_\xi P-\frac{1}{\Sigma_0}\partial_\xi P_0
-c\partial_\xi\frac{2c}{\gamma-1}\pm cu\frac{\partial_\xi\Sigma}{\Sigma}
\mp u\partial_\xi\frac{2c}{\gamma-1}\right]\nonumber\\
+\frac{2\Omega}{\Delta \Omega} v +\frac{v^2}{\Delta \Omega r}\mp 
\frac{c u}{\Delta \Omega r}\mp \frac{c v}{\Delta \Omega r}
\partial_\phi\ln\Sigma
\mp \frac{c}{\Delta \Omega r}\partial_\phi v,
\end{eqnarray}
where
\begin{eqnarray}
R_\pm=u\pm \frac{2c}{\gamma-1},
\end{eqnarray}
and it is always assumed that $c=c(r)=c(\xi)$.

As noted in GR01, the characteristic $C_+$ {\it crosses} 
the wake profile 
while the characteristic $C_-$ {\it follows} it, that is $C_-$ is always  
in the perturbed region while $C_+$ mostly passes through the 
unperturbed fluid and only for short periods of time enters the perturbed
region. One can assume the 
changes in $R_+$ due to these crossings to be small and 
take  $R_+$ to be constant (we will later comment on the effect of the 
multiple crossings) and equal to its value in the unperturbed region. 
There,
$R_+=2c_0/(\gamma-1)$ everywhere.  
Then, from the conservation of $R_+$ one obtains
 that $u=2(c_0-c)/(\gamma-1)$ and
$R_-=2(c_0-2c)/(\gamma-1)$. 

Using these results it is demonstrated in Appendix that the 
equation 
of evolution of $R_-$ can be reduced to an inviscid Burger's equation
\begin{equation}
\partial_{t}\chi-\chi\partial_{\eta}\chi=0,\label{burg}
\end{equation}
where 
\begin{eqnarray}
\chi\equiv\frac{\gamma+1}{2}\frac{\Sigma-\Sigma_0}{\Sigma_0}g(r),\label{chi}\\
t\equiv -\frac{r_p}{l_p}\int\limits^r_{r_p}
\frac{\Omega(r^\prime)-\Omega_p}{c_0(r^\prime) g(r^\prime)}dr^\prime,
\label{t_fin}\\
\eta\equiv\frac{r_p}{l_p}\left[\phi+\mbox{sign}
(r-r_p)\int\limits^r_{r_p}\frac{\Omega(r^\prime)-\Omega_p}
{c_0(r^\prime)}dr^\prime\right],\label{eta_fin}\\~~~\mbox{and}~~~
g(r)\equiv\frac{2^{1/4}}{r_p c_p \Sigma_p^{1/2}}\left(
\frac{r \Sigma_0 c_0^3}{|\Omega-\Omega_p|}\right)^{1/2}.\label{g}
\end{eqnarray}
Here $l_p=c_p/|2A(r_p)|=(2/3)h_p$ 
is a Mach-$1$ length (distance from the planet
where the Keplerian shear makes fluid velocity equal to $c_p$). 

In the immediate vicinity of the planet (but still several
$h_p$ from it) definitions (\ref{chi})-(\ref{g}) reduce to
\begin{eqnarray}
\chi\to \frac{\gamma+1}{2^{3/4}}\bigg|
\frac{x}{l_p}\bigg|^{-1/2}\frac{\Sigma-\Sigma_0}{\Sigma_0},~~~
t\to \frac{2^{3/4}}{5}\bigg|
\frac{x}{l_p}\bigg|^{5/2},~~~\eta\to\frac{y}{l_p}+\frac{x^2}{2 l_p^2}
\mbox{sign}(x),
\label{local}
\end{eqnarray}
where $x=r-r_p$ and $y=r_p\phi$. These equations coincide with 
analogous expressions in GR01.

The tidal perturbation launched by the planet follows a nearly parabolic path 
in the immediate vicinity of the planet where the shear can be assumed
constant. Further from the perturber where the shear is no longer uniform,
the density wave has a spiral shape. Its pattern is described 
by the equation 
\begin{equation}
\phi=\phi_0-\mbox{sign}(r-r_p)\int\limits_{r_p}^r
\frac{\Omega(r^\prime)-\Omega_p}
{c_0(r^\prime)}dr^\prime\label{spiral}
\end{equation}
Depending on the conditions
in the disk this spiral can wind up several times around the center before
the wave damps. One can see from (\ref{spiral}) that the perturbation is indeed
tightly wound if $|r-r_p|\gg h_p$ (this could, however, be violated further
away from the planet
for some profiles of $\Sigma_0$ and $c_0$, see \S \ref{pow-law-disks}).

Now, if we disregard the nonlinear evolution entirely, it follows 
from equation (\ref{burg}) that $\chi=f(\eta)$ [$f(x)$ 
is an arbitrary
function of the argument $x$], or
\begin{equation}
(\Sigma-\Sigma_0)_{lin}=\frac{\Sigma_0}{c_0 g(r)}f(\eta)=\left[
\frac{\Sigma_0 (\Omega-\Omega_p)}{r c_0^3}\right]^{1/2}
f\left(\phi+\mbox{sign}(r-r_p)
\int\limits_{r_p}^r\frac{\Omega(r^\prime)-\Omega_p}
{c_0(r^\prime)}dr^\prime\right).
\end{equation}
This means that in the linear regime, the waveform  
 propagates along the wake (whose location is given by
the condition $\eta=const$) and its amplitude scales with the distance
from the planet in complete
agreement with  equation
(\ref{scaling}).

\section{Nonlinear evolution of the wake.}\label{wakeprop}

Burger's equation (\ref{burg}) is probably the simplest partial 
differential equation able to exhibit a shock formation phenomenon.
For the reasons outlined in \S \ref{gencons} the profile of the density 
wave produced by the linear generation mechanism is constantly distorted 
in the course of its propagation away from the planet, so that finally 
the waveform
breaks to become double-valued. This implies that a shock must be formed
at this point.
The distance which the density wave travels before it shocks
depends on the initial shape and the amplitude of the wake. 

To study shock formation and propagation quantitatively
we need to solve 
equation (\ref{burg})
with the initial condition given by the solution of the linear wake generation 
problem. As we have mentioned in \S \ref{basder},
since all our variables reduce to the corresponding variables
of GR01 in the limit $r\to r_p$, the whole linear generation problem reduces
to the one solved before --- wake excitation in the shearing sheet 
approximation. 
Thus, we might use the solution for the wake shape calculated in GR01  
as the initial condition in our more general case: 
\begin{equation}
\chi(M_p,t=t_0,\eta)=\frac{M_p}{M_1}f_0(\eta)\label{bound}
\end{equation} 
[$M_1$ is defined in equation (\ref{M1})].
Here the initial profile $\chi(M_p,t=t_0,\eta)$ is taken not at $t=0$ but
at some $t_0>0$ because in the linear regime
 wake has to propagate some distance 
from the planet before it fully forms. Thus $t_0$
is a matching boundary where one switches from 
linear to weakly nonlinear regime,
 and following GR01 we take $t_0\approx 1.89$ corresponding to
the distance from the planet $|x_0|=2l_p=(4/3)h_p$. One can usually
neglect $t_0$ once the wave has travelled several $h_p$ away from the planet.

\begin{figure}
\vspace{10cm}
\includegraphics{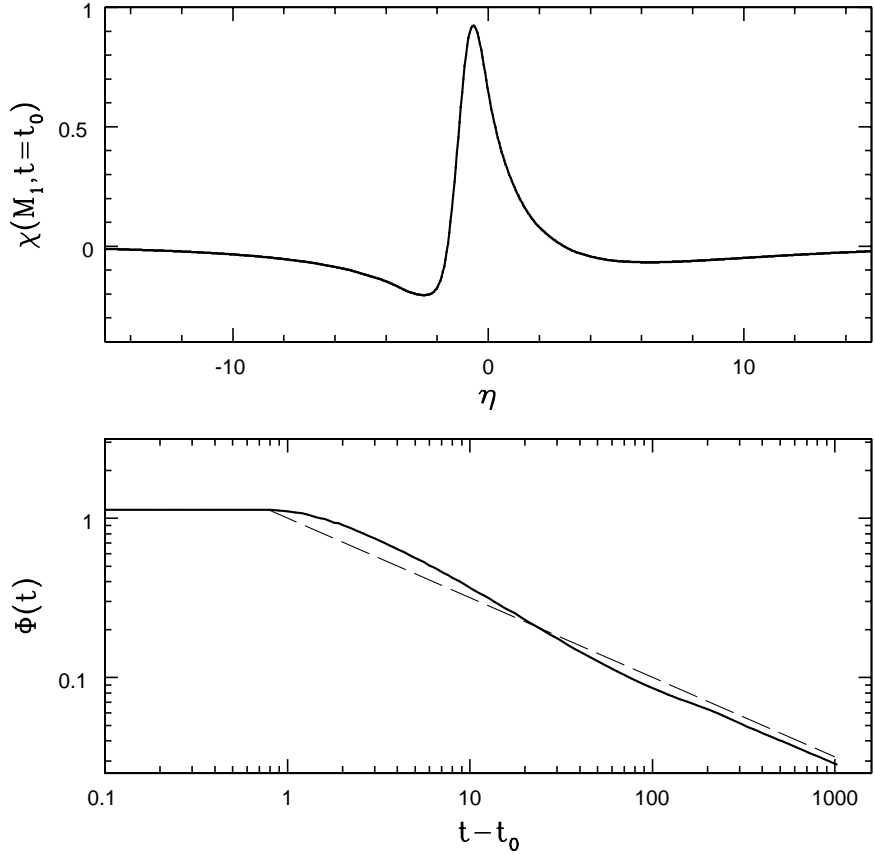}
\caption{
({\it top}) Shape of the wake profile at the moment $t=t_0\approx 1.89$ 
calculated in the linear theory. This profile is used as initial
condition for the calculation of the nonlinear evolution of the wake.
({\it bottom}) Behavior of the dimensionless angular momentum flux $\Phi(t)$
carried by the waves in the presence of shock damping. The 
calculation is done for
$M_p=M_1$. The dashed line has a slope of $t^{-1/2}$ and is drawn 
to illustrate 
the asymptotic behavior of $\Phi(t)$.
\label{fig:wake}}
\end{figure}

The function $f_0(\eta)$ represents
 the shape of the initial profile at the moment
$t=t_0$ for $M_p=M_1$ [see equation (\ref{M1})]. It was calculated by GR01 
in the shearing sheet approximation
and is reproduced for convenience in Fig. \ref{fig:wake}a.
The factor $M_p/M_1$ rescales the amplitude of the wake for an arbitrary
mass of the planet.

In terms of the variables $\chi, \eta, t$ our nonlinear 
problem (\ref{burg}) is identical to the one
studied in the shearing sheet approximation [including the initial condition
(\ref{bound})]. This means that the result of GR01 
for $\chi(M_1,t,\eta)$ in terms of variables
$\chi,\eta,t$ is also a solution of our more general problem in 
terms of these variables.
One needs only to express them   
in terms of $r$ and $\phi$ using equations (\ref{chi})-(\ref{g}) and rescale
to an arbitrary mass $M_p$ in the following way:
suppose that $\chi_1(t-t_0,\eta)\equiv \chi(M_1,t-t_0,\eta)$ is a solution 
of the equation (\ref{burg})
with the boundary condition (\ref{bound}) and $M_p=M_1$ 
(calculated in GR01). Then one can easily see
that the solution for an arbitrary $M_p$ can be written as
\begin{equation}
\chi(M_p,t-t_0,\eta)=\frac{M_p}{M_1}\chi_1\left(\frac{M_p}{M_1}(t-t_0),
\eta\right).\label{gensol}
\end{equation}

This reduction to the previously studied case  
is a very convenient feature of the analysis because it allows one to use all 
results previously obtained by GR01  in our more general setting.
For instance, it was found before that 
the shock develops at 
\begin{equation}
t_{sh}=t_0+0.79\frac{M_1}{M_p}\label{pointshck}
\end{equation}
for the initial profile
given by the function $f_0(\eta)$ (see Fig. \ref{fig:wake}a).
Based on what we have previously
said, the shock has to form at the 
same value of $t_{sh}$ in our global case, 
and the corresponding radial distance from the planet
$r_{sh}-r_p$ at which this happens can be found using equations (\ref{t_fin}) 
\& (\ref{g}).
For a fixed $M_p$ this procedure leads to the condition
(\ref{shck}) found before on the basis of qualitative 
considerations.

After the wave shocks, it starts transferring its angular momentum to
the disk mean flow which leads to the damping of the wave amplitude. 
The initial profile shown in Fig. \ref{fig:wake}a has both positive 
and negative
parts, thus it is destined to evolve asymptotically 
into an ``N-wave'' profile (Landau \& Lifshitz 1959; Whitham 1974). 
In this regime the amplitude of $\chi$ decays as
$ t^{-1/2}$, whereas it was constant near 
 the planet before forming a shock.
The width of the spiral perturbation, which was
$\sim h_p$ initially,  grows as 
$t^{1/2}$ in the ``N-wave'' stage.

\subsection{Angular momentum flux.}\label{ang-mom}

Summing up all the angular momentum flux contributions given by 
equation (\ref{fjm-sig}) and 
applying Parseval's theorem
one can show that the total
angular momentum flux of the density wave $f_J$ is given by:
\begin{equation}
f_J(r)=\frac{c_0^3(r) r}{[\Omega(r)-\Omega_p] 
\Sigma_0(r)}\int\limits_{-\pi}^\pi
(\Sigma-\Sigma_0)^2d\phi.
\end{equation}

Using definitions 
(\ref{chi}) and (\ref{g}) we obtain that 
in terms of variables $t$ and $\eta$
\begin{eqnarray}
f_J(t)=\frac{2^{3/2}c_p^3 r_p\Sigma_p}{(\gamma+1)^2|2A(r_p)|}\Phi(M_p,t),
~~~~\mbox{where}\nonumber\\
\Phi(M_p,t)=\int\chi^2(M_p,t,\eta)d\eta=
\left(\frac{M_p}{M_1}\right)^2\Phi\left(M_1,\frac{M_p}{M_1}t\right),
\label{phdef}
\end{eqnarray}
exactly like in GR01 (we neglected $t_0$ here). 
From equation (\ref{burg}) one can easily see that without shocks
$\Phi(M_p,t)= const$ and angular momentum flux is conserved.

After the shock forms $\Phi$ starts to decrease. Asymptotically, 
in the ``N-wave'' regime
 it falls like $\Phi(M_p,t)\propto t^{-1/2}$. The behavior of $\Phi$
as a function of $t$ for $M_p=M_1$ was calculated in GR01 and 
is shown in Fig. \ref{fig:wake}b. 
What happens with the angular momentum flux of the wave in the 
disk depends on how $t$ is related to the distance travelled by the wave. 
If one assumes a uniform shear, then according to equation (\ref{local})
$t\propto |r-r_p|^{5/2}$
and at infinity the wave damps completely. Real disks are 
different --- they are always finite,
dissipation can be not complete, 
and the damping pattern is asymmetric between the inner and outer disks.
We will demonstrate later that for some background
surface density and sound speed profiles $t(r=\infty)$ or $t(r=0)$ 
are finite. This means that a density wave that has managed to propagate to 
the outermost parts of the disk, or to its center, still carries some 
undamped angular
momentum flux. 
It could even happen that $t(r=\infty)<t_{sh}$ or $t(r=0)<t_{sh}$,
meaning that the wave does not shock before it reaches the outer or inner 
edge of the disk. The wave could then be reflected and part of its angular 
momentum could 
be carried back to the planet. This might have important consequences
for planetary migration and we will dwell upon this 
more in \S \ref{disc}.

\subsection{Power-law disks}\label{pow-law-disks}

We now consider the case of power-law disks, i.e. $\Sigma_0$ and $c_0$
are assumed to be some powers (usually negative) of $r$, 
to illustrate the results 
obtained before and test some of our assumptions. 
We consider a Keplerian disk with
$U_\star\propto r^{-1}$ and take $\Sigma_0=\Sigma_p(r/r_p)^{-\delta}$
and $c_0=c_p(r/r_p)^{-\nu}$, $\delta>0, \nu>0$.

Using definitions (\ref{chi}), (\ref{t_fin}), and (\ref{g})
we find that
\begin{eqnarray}
t=\frac{3}{2^{5/4}}
\left(\frac{r_p}{h_p}\right)^{5/2}\Bigg|
\int\limits_1^{r/r_p}|s^{3/2}-1|^{3/2}s^{
(5\nu+\delta)/2-11/4}ds \Bigg|,\label{kept}
\end{eqnarray}
and the wake equation
\begin{eqnarray}
\phi=\phi_0-\mbox{sign}(r-r_p)\frac{r_p}{h_p}\left\{
\frac{(r/r_p)^{\nu-1/2}}{\nu-1/2}-
\frac{(r/r_p)^{\nu+1}}{\nu+1}-
\frac{3}{(2\nu-1)(\nu+1)}\right\}.\label{kepphi}
\end{eqnarray}

\begin{figure}
\vspace{8cm}
\includegraphics{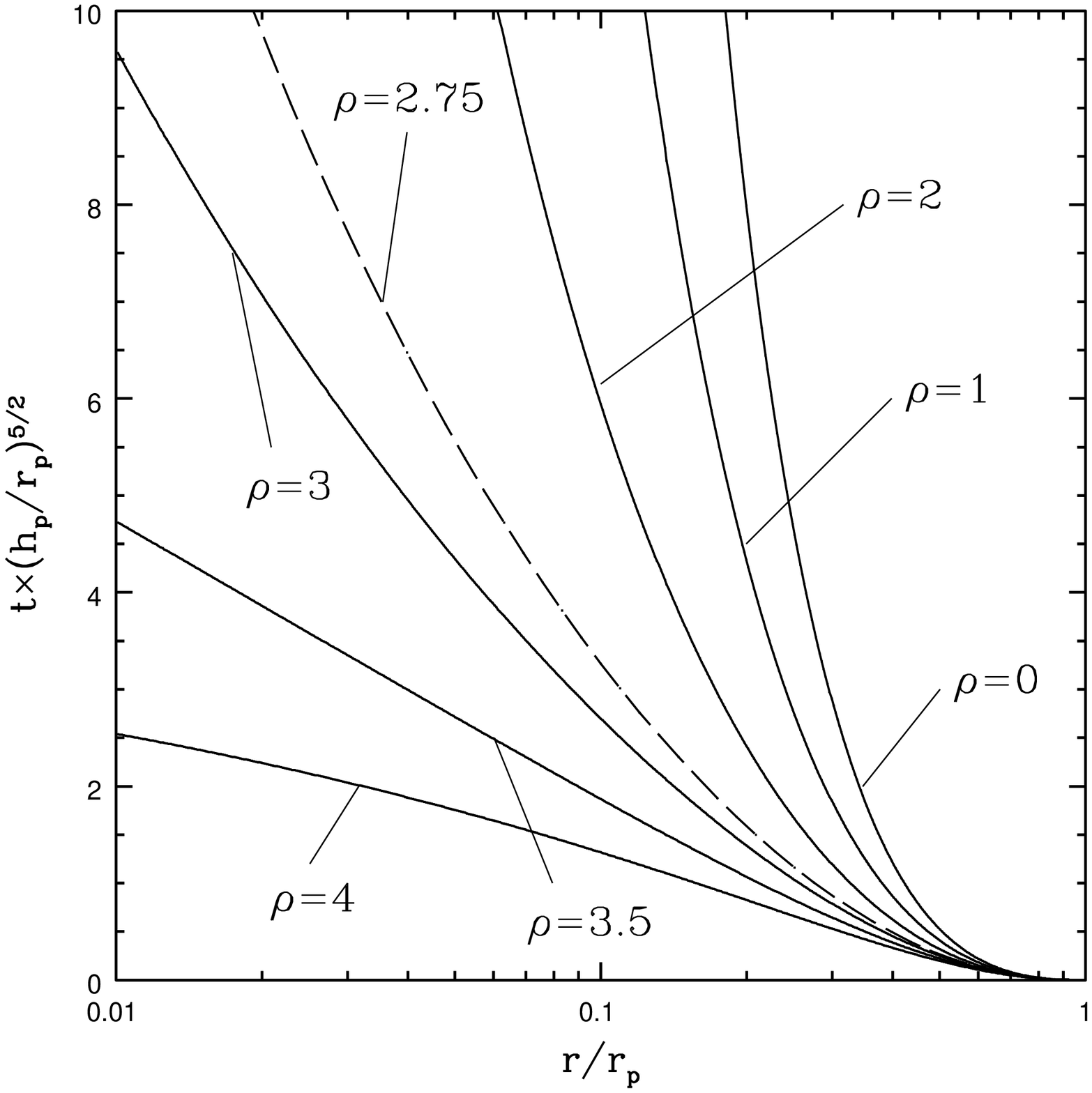}
\includegraphics{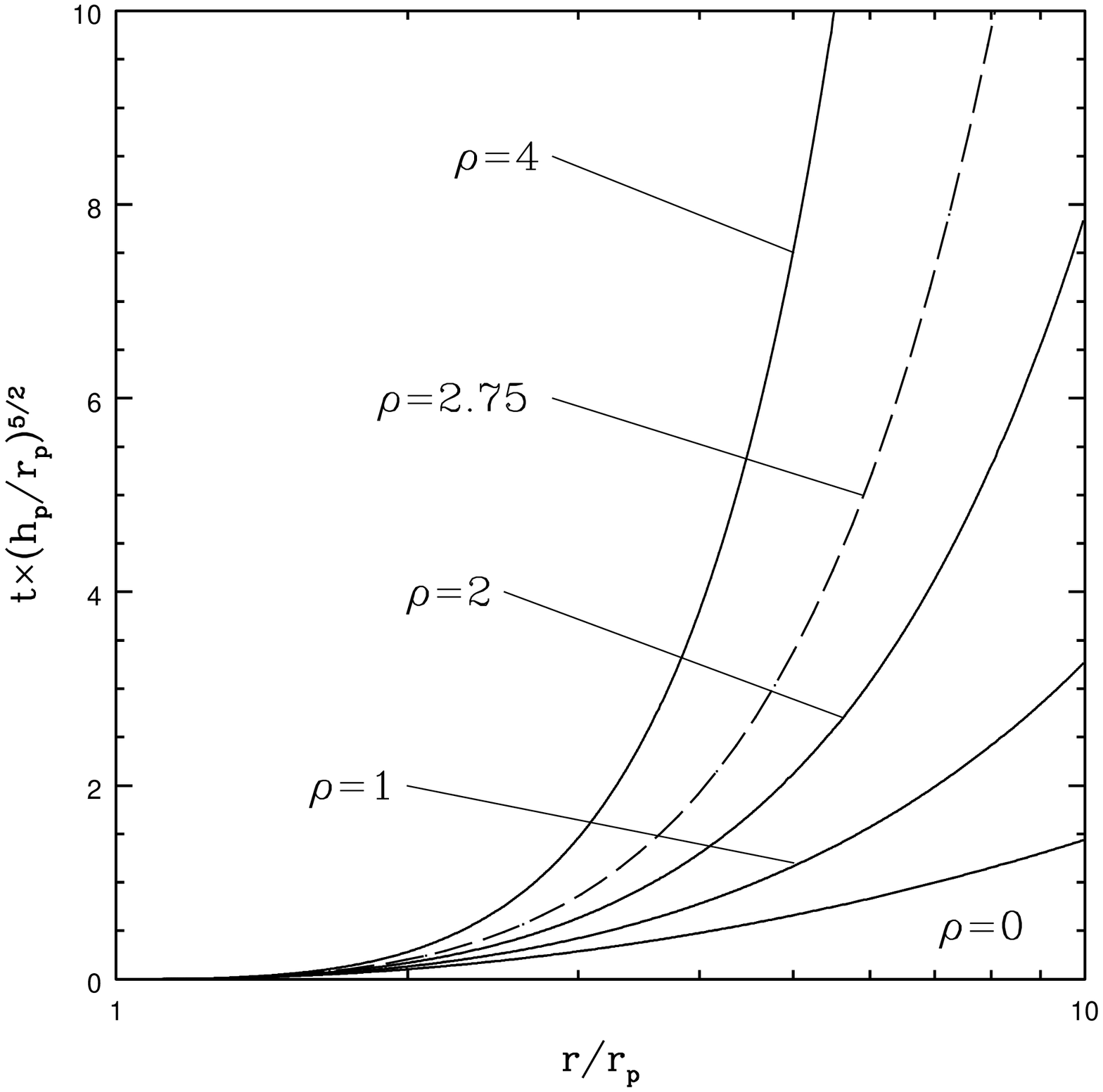}
\caption{Dependence of the dimensionless variable $t$ [multiplied by 
$(h_p/r_p)^{5/2}$] upon the distance
from the central body $r$ for several values of the power-law index 
$\rho=5\nu+\delta$ (see the beginning
of \S \ref{pow-law-disks}). 
Left panel is for $r<r_p$, right one is for $r>r_p$.
Curves are labelled by the corresponding values
of $\rho$. Notice a universal behavior 
of $t(r)$ near $r=r_p$, where the shearing sheet approximation is valid.
In the inner part of the disk 
$t(r)$ is increasing to infinity  as $r\to 0$ 
(which implies a complete dissipation 
of the shock) only for $\rho<\rho_{cr}=7/2$. In the outer disk wave
shocks and damps for all values of $\rho$ as $r\to\infty$.
\label{Fig1}}
\end{figure}

Using equation (\ref{kept})
let us consider first the inner ($r<r_p$) 
part of the disk. One can easily see that $t\to \infty$
as $r\to 0$ if $\rho=5\nu+\delta<\rho_{cr}=7/2$. This means that the 
density waves launched by the planet 
at $r_p$ in the inner disk  span 
the whole range of $t$ and thus damp completely upon reaching the center.
Of course, real disks always have an inner cavity, presumably formed
by the protostellar magnetospheric activity, where the gas is absent, so we
only have to consider wave propagation up to this inner edge. Depending on 
the location of the planet, this inner disk boundary may be very close
to the center so that we will consider wave propagation 
down to $r=0$.   

For $\rho>\rho_{cr}$
however, $r=0$ corresponds to a finite $t$, and if
$t(0)<t_{sh}$ the wave does not shock at all as $r$ decreases to $0$.
Indeed, for $\rho=5\nu+\delta>\rho_{cr}=7/2$ one obtains that
\begin{equation}
t(0)=\frac{1}{2^{1/4}}
\left(\frac{r_p}{h_p}\right)^{5/2}B\left(\frac{5\nu+\delta}{3}-\frac{7}{6},
\frac{5}{2}\right),
\end{equation}
where $B(\alpha,\beta)$ is a beta-function. Using equation (\ref{pointshck}) 
and neglecting $t_0$ in it we get that if 
\begin{equation}
M_p<M_x\equiv M_1\times 0.94\left(\frac{h_p}{r_p}\right)^{5/2}\left[B\left(\frac{5\nu+\delta}{3}-\frac{7}{6},\frac{5}{2}\right)\right]^{-1},\label{shckcond}
\end{equation}
then the wave does not shock on the first passage to the disk center. For 
this to happen $M_p$ should be sufficiently small: assuming that
$B(\alpha,\beta)\sim 1,~ r_p/h_p \gg 1$ one needs 
$M_p\la M_1(h_p/r_p)^{5/2}$
for this to occur. If $\rho>\rho_{cr}$ but condition (\ref{shckcond})
is not fulfilled (if the planet is massive enough), the wave shocks as it 
travels towards the disk center, but
it does not damp completely upon reaching it. Part of the angular momentum 
flux could be transferred back to the planet in this case.

\begin{center}
\begin{deluxetable}{ l l c }
\tablecolumns{3}
\tablewidth{0pc}
\tablecaption{Wave behavior in the infinite disk for different 
values of $\rho>0$. 
\label{table}}
\tablehead{
\colhead{Outcome of the wave propagation}&
\colhead{Inner disk}&
\colhead{Outer disk}
	}
\startdata
Wave shocks, damps completely  & $0<\rho\le\rho_{cr}$, any $M_p$ & any $\rho>0$, any $M_p$ \\
Wave shocks, but damps not completely & $\rho>\rho_{cr}$, $M_p\ge M_x$ & --- \\
Wave does not shock and does not damp & $\rho>\rho_{cr}$, $M_p< M_x$ & --- \\
\enddata
\end{deluxetable}
\end{center}

One can easily see that in the outer disk this problem never arises:
for $\delta>0, \nu>0$ the wave always shocks as it propagates outwards, 
if we forget about the outer boundary of the disk.
In Table 1 we summarize final outcomes of wave propagation in the inner
and outer parts of the disk for different values of $\rho$.
In Fig. \ref{Fig1} 
we plot the behavior of $t(r)$
for different values of $\rho$ in the inner and outer parts of 
the disk using equation (\ref{kept}). 
In the minimum mass Solar nebula (MMSN) $\delta=3/2$ and $\nu=1/4$
(Hayashi 1981) meaning that $\rho=11/4<\rho_{cr}$. Thus, in MMSN  
tidal waves always shock and are damped on both sides of the disk.

\begin{figure}
\vspace{8cm}
\includegraphics{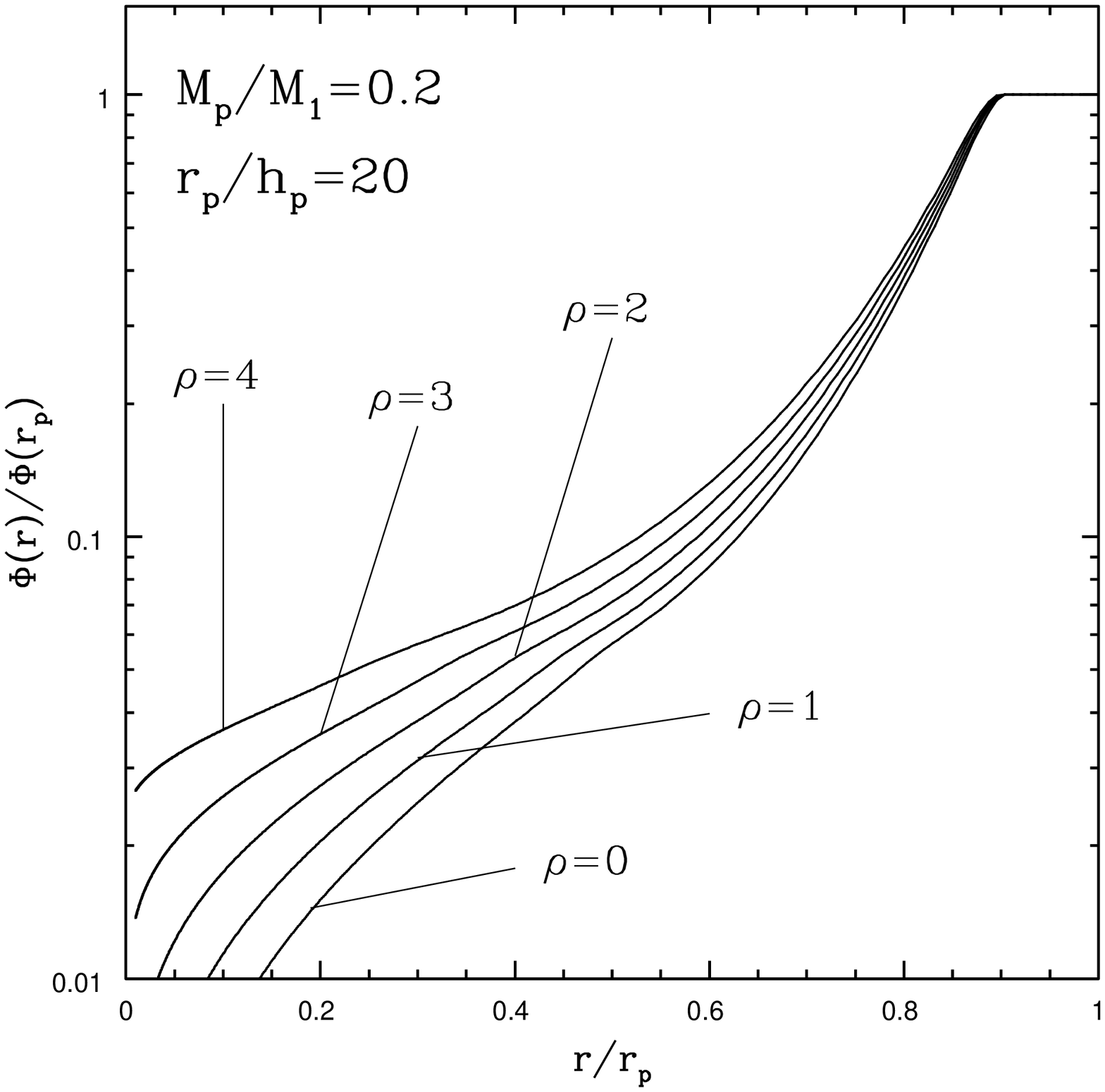}
\includegraphics{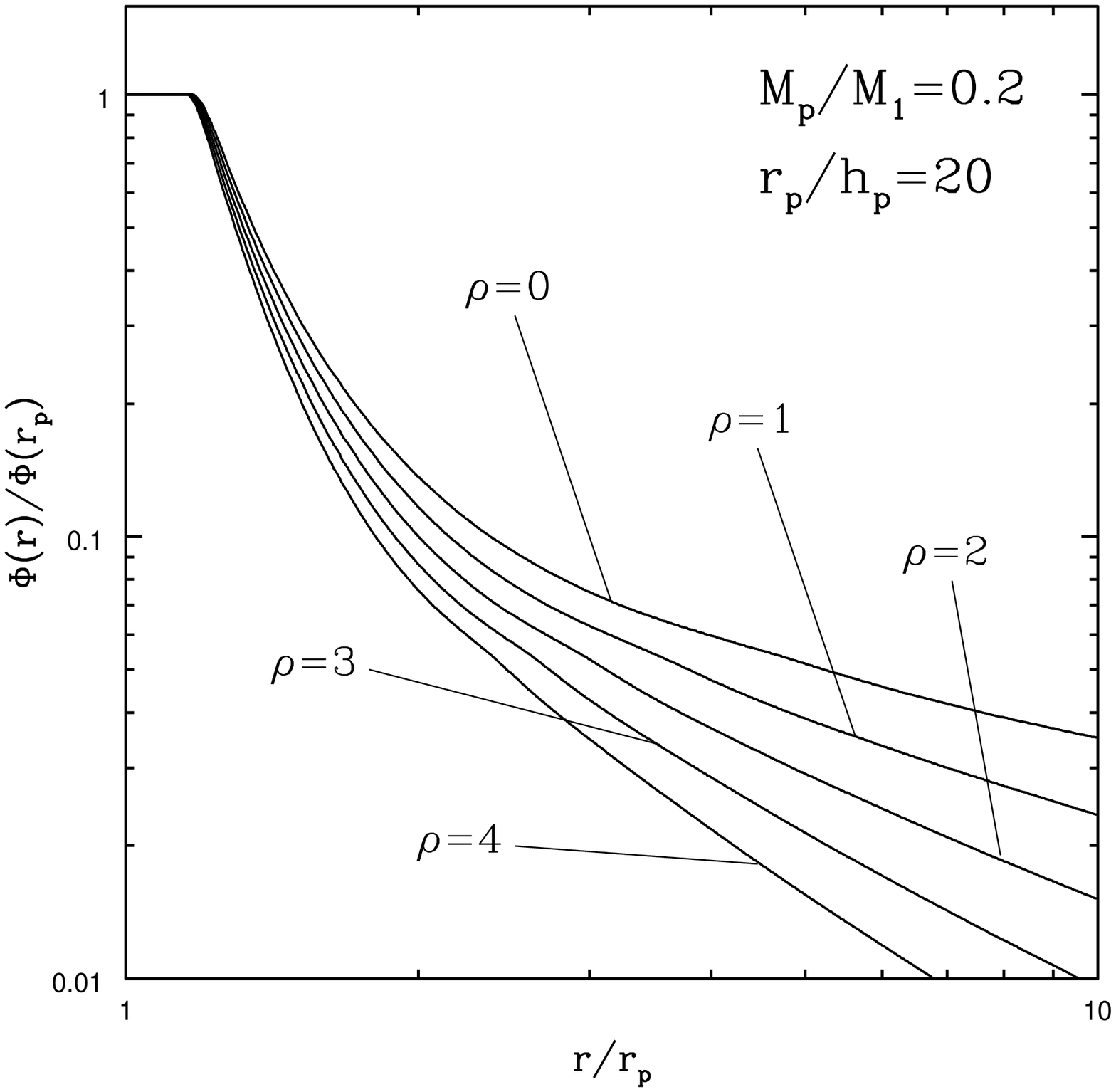}
\caption{Dependence of the dimensionless 
angular momentum flux of the planet-generated 
density wave $\Phi(r)$ upon the distance from the central body $r$
for several values of the power-law index $\rho=5\nu+\delta$. 
Left panel is for the inner part of the disk, 
right is for the outer one (notice that on the 
left panel horizontal scale is linear). 
Calculation assumes that $M_p=0.2~M_1$ [see equation (\ref{M1})] and ratio
$r_p/h_p=20$. Different curves are labelled by the corresponding values
of $\rho$. Notice that in the case $\rho=4>\rho_{cr}$ angular momentum flux
is nonzero at the disk center. 
\label{Fig2}}
\end{figure}

It is interesting to see how the shock damping is distributed in the disk.
Using the asymptotic behavior of $\Phi(t)$ for large $t$ we find that 
\begin{equation}
f_J\propto \left(\frac{r}{r_p}\right)^{\frac{7-2\rho}{8}},\label{scalefj}
\end{equation}
as $t\to 0$. As $\rho$ varies from $0$ to $\rho_{cr}$ the power law index here
changes from $7/8$ to $0$ (it is equal to $3/16$ for MMSN), 
which implies that for some $\Sigma_0$
and $c_0$ profiles quite a lot of the residual angular momentum is 
transferred to the disk close to its center. This might significantly 
enhance the accretion rate there.
In the outer disk
\begin{equation}
f_J\propto \left(\frac{r}{r_p}\right)^{\frac{\rho+1}{2}},\label{scalefjout}
\end{equation}
for $r\to\infty$.
For shallow profiles of $\Sigma_0$ and $c_0$ the
nonlocality of the damping could be important in the outer disk too.
The behavior of the
dimensionless angular momentum flux 
$\Phi(r)$ is shown in Fig. \ref{Fig2} for different values of $\rho$. 

Let us now test the validity of some of our 
assumptions which were used in the analysis presented in \S \ref{basder}.
The tight-winding approximation is one of them. From the equation (\ref{spiral})
we find that 
\begin{equation}
\tan\theta=r\frac{\Omega(r)-\Omega_p}{c_0(r)}=
\frac{r_p}{h_p}\left(\frac{r}{r_p}\right)^{\nu+1}
\left[\left(\frac{r_p}{r}\right)^{3/2}-1\right],\label{tight}
\end{equation}
where $\theta$ is an angle between the radial direction and the 
tangent of the spiral, and the last equality in (\ref{tight})
is for a Keplerian power-law disk. In the immediate vicinity of the planet,
for $|r-r_p|\sim h_p$, this angle is small and spiral pattern is
not tightly wound. But this is the region of the wake generation where
the free nonlinear propagation approach is not applicable anyway. 
After travelling several $h_p$ from the planet, the 
 wake becomes tightly wrapped
by the Keplerian shear if the disk thickness is small
($r_p/h_p\gg 1$).
In the outer disk asymptotically $\theta\to \pi/2$
as $r\to\infty$ (spiral winds up). 
This happens because in the frame rotating with the
planet outer parts of the disk move with the angular speed $-\Omega_p$
giving rise to a large linear velocity, while $c_0$ decreases with 
growing $r$.
Inside the planet's location $\tan\theta\approx(r_p/h_p)(r/r_p)^{\nu-1/2}$ 
as $r\to 0$. Thus, for $\nu>1/2$ the spiral pattern unwinds in the inner
regions of the disk
and tight-winding approximation might become inapplicable. 
However, if $r_p/h_p$ is large enough, wave could 
still reach the inner disk edge before this effect becomes important
(in fact, spiral pattern unwinds only if $h\sim r$ which is 
usually not the case 
in the inner part of the disk).

Geometrical effects may also be of some importance. As we mentioned before,
in the asymptotic regime the wake width increases as $t^{1/2}$.
At the same time, if the 
pattern of the wake winds up, the distance between the
consecutive wavecrests (at a fixed polar angle $\phi$) decreases. At some 
point the ``rear'' shock front of the ``N-wave'' profile comes so close 
to the ``forward'' shock front of the profile lagging by $2\pi$ in $\phi$
that our approximation of almost constant Riemann invariant $R_+$ 
becomes poor because the change of $R_+$ during the shock crossing 
gets comparable to the change of $R_-$ following the shock 
 (see the discussion in \S \ref{basder}). 
  
From Fig. 2 of GR01 one can find that the width of the 
shock in the $\eta$ coordinate in the asymptotic region $t\to\infty$ 
is $\Delta \eta\approx 2.3 t^{1/2}$. 
Obviously the ``front touching'' phenomenon occurs when 
$(l_p/r_p)\Delta \eta\ga 
2\pi$ [see equations (\ref{varch}) and (\ref{renorm})], that is when 
$t(h_p/r_p)^{5/2}\ga 17 (h_p/r_p)^{1/2}$. This particular form 
of the condition is used because it 
allows direct comparison with Fig. \ref{Fig1}. One can see 
from this Figure that for 
$r_p/h_p=20$ and any $\rho>0$ one 
can propagate as far as $r/r_p\approx 0.3$ in the inner disk 
and $r/r_p\ga 4$ in the outer,
and still not encounter this ``front touching''. In colder disks 
with higher $r_p/h_p$ this limitation becomes more stringent but
even in this case there is a significant region of applicability
of our analysis not too far from the planet 
where $R_+$ could be assumed approximately constant. 

We also need to mention that our treatment of the nonlinear evolution
 essentially neglected linear dispersion of the wave profile. 
Near the planet wave dispersion is strongest but  
it is properly taken there into
account in the linear calculations of GR01. Further away from the planet,
according to equation (\ref{disp}),
dispersion 
rapidly becomes less important in comparison with nonlinear steepening
(if $M_p$ is not very small) and can be ignored. 

\section{Discussion and applications.}\label{disc}

It is quite possible that in the presence of 
vertical temperature  gradients 
the channeling of the wave action into the 
vertical direction can damp density waves in the disk atmosphere
more efficiently than we have found here (Lin et al. 1990;
Lubow \& Ogilvie 1998).
It seems reasonable however that in passive disks heated by their central 
stars thermal stratification in the $z$-direction must be small because
of their high optical depth. This strongly diminishes the 
wave action channeling into the disk atmosphere (Ogilvie \& Lubow 1999)
and leads to almost two-dimensional 
picture of the wave propagation in the disk 
supporting the validity of our consideration. 

Throughout our analysis we assumed the wave nonlinearity to be weak,
meaning that $(\Sigma-\Sigma_0)/\Sigma_0$ is small. From equation (\ref{bound})
we see that if $M_p\ga M_1$ the wake is nonlinear from the very beginning
and it shocks immediately
[see equation (\ref{pointshck})]. This means that the separation of the disk
into two distinct regions where one can neglect either planetary torques
or nonlinearity of the wave does not exist.
Also for $M_p\ga M_1$ a gap in the disk can form around the planet 
(Lin \& Papaloizou 1993).
Thus, as we have mentioned before in \S \ref{basder},
our analysis is applicable only for small-mass planets:
$M_p\la M_1$.

We have seen in \S \ref{pow-law-disks} that wave damping can be a 
nonlocal process and that part of the wave action could reach 
the disk edge undamped.
This incomplete damping  is important for the question
of the planetary migration. The migration speed and direction depend 
sensitively on a delicate balance between the amounts of the 
angular momentum 
which the planet deposits into the inner and outer parts of the disk. 
If the density waves  dissipate completely and transfer all their
angular momentum to the disk fluid then the
only difference in torques acting
on both sides of the disk is due to the 
surface density and temperature gradients
in the disk, and to asymmetries in the locations of the inner and outer 
Lindblad resonances
(Ward 1986). The amount
of the resultant torque which leads to the orbital evolution of the 
planet is only $\sim h_p/r_p$ compared to the magnitude of  
one-sided torque (GT80). Interaction with the outer part
of the disk is usually {\it stronger} than with the inner one, leading to an
{\it inward} migration (Ward 1986). 
 
Let us now assume that tidal perturbations are not damped completely upon 
reaching the disk edge and are reflected from it. The remaining 
waves will be dissipated  on the way back to the planet
but some of them might survive and  interact gravitationally with the planet,
returning to it some of the initially launched angular momentum. 
If this effect is able to return to the planet about $h_p/r_p$ of the 
one-sided angular momentum then the migration could be strongly modified
[Tanaka \& Ward (2000) studied a similar effect caused by 
asymmetries in wave damping].
Consider, for instance,  
a planet sitting close to the outer edge of the disk, but still 
several $h_p$ from it (otherwise strong asymmetries in the torque
will be produced when the wake is still forming).
A tidal wave launched in the inner disk might be completely dissipated
because of the large distance it has to travel in the disk. 
At the same time the wave in the outer disk might not shock at all before 
being reflected from the disk edge and can
bring a significant amount of the angular
momentum back to the planet. Thus, interaction with the outer part of the 
disk is now {\it weaker} than with the inner one and migration will
change its direction -- the planet will move outwards. 
In the same way one could show
that planets near the inner edge of the disk tend to move inwards
faster if the waves are incompletely damped in the inner part of the disk.
One can roughly describe this process by saying that the planet tends to be 
pushed out from the disk towards its closest boundary. 
The same picture will hold for wave reflection off the edges of gaps formed 
by giant planets.
These conclusions depend on a lot of assumptions, such as the details of 
the reflection process and gravitational interaction of the reflected wave
with the planet, which certainly deserve further study. Whether this 
process is an 
interesting issue for the question of planet formation and survival 
in the course of migration depends on the relevant timescales. Nevertheless, 
incomplete damping of the density waves introduces additional degree of 
freedom on which the migration process depends and
which could be important in some systems.

Deposition of the wave angular momentum into the disk fluid leads to the 
evolution of the disk itself (Larson 1989). Spruit (1987) and Larson (1990)
found that the action of shocked density waves 
is equivalent to an appreciable viscosity with corresponding dimensionless 
$\alpha$-parameter (Shakura \& Sunyaev 1973) reaching $\sim 10^{-4}-10^{-3}$. 
However they did not specify the source of the tidal perturbation, a deficiency
remedied in GR01. There it was demonstrated that if all the solids
in the disk were deposited into a population of Earth-sized objects 
then, again, an effective viscosity $\alpha\sim 10^{-4}-10^{-3}$
is produced. 

Our results allow one to study the global evolution of the disk 
affected by all the planets present in it.
The theory of the time-dependent accretion disks
(Pringle 1981) states that the mass accretion rate at each point in the disk
is uniquely 
determined by the divergence of the angular momentum flux carried by the
waves:
\begin{equation}
\dot M=\left[\frac{\partial }{\partial r}(r^2\Omega)\right]^{-1}
\frac{\partial f_J}{\partial r}.\label{dotm}
\end{equation}
Since $f_J$ depends on the distance travelled by the 
wave in a complex way [see equation (\ref{phdef})], one should expect 
$\dot M$ produced by a single planet to be a function of $r$. 
Using (\ref{phdef}), (\ref{kept}), \& (\ref{dotm}) we can calculate the
accretion rate in the disk at a distance $r$ from the center
produced by a single planet  located at $r=r_p$:
\begin{eqnarray}
\dot M(r)=\mbox{sign}(r-r_p)\frac{2^{9/4}}{(\gamma+1)^2}
\left(\frac{h_p}{r_p}\right)^{1/2}\Omega(r_p)\Sigma_0(r_p)r_p^2
\left(\frac{M_p}{M_1}\right)^{3}\nonumber\\
\times\Bigg| \left(\frac{r}{r_p}\right)^{3/2}-1 \Bigg|^{3/2}
\left(\frac{r}{r_p}\right)^{\rho/2-9/4}\Phi^\prime\left(
M_1,\frac{M_p}{M_1}t(r/r_p)\right).\label{tork}
\end{eqnarray}
Here $\Phi^\prime(x)\equiv d\Phi(x)/dx$.
Note that because $\Phi^\prime<0$, the accretion rate 
is positive inside of the planet (inflow) and
negative outside of it (outflow) --  the perturber tries to repel 
the surrounding gas.

\begin{figure}
\vspace{10cm}
\includegraphics{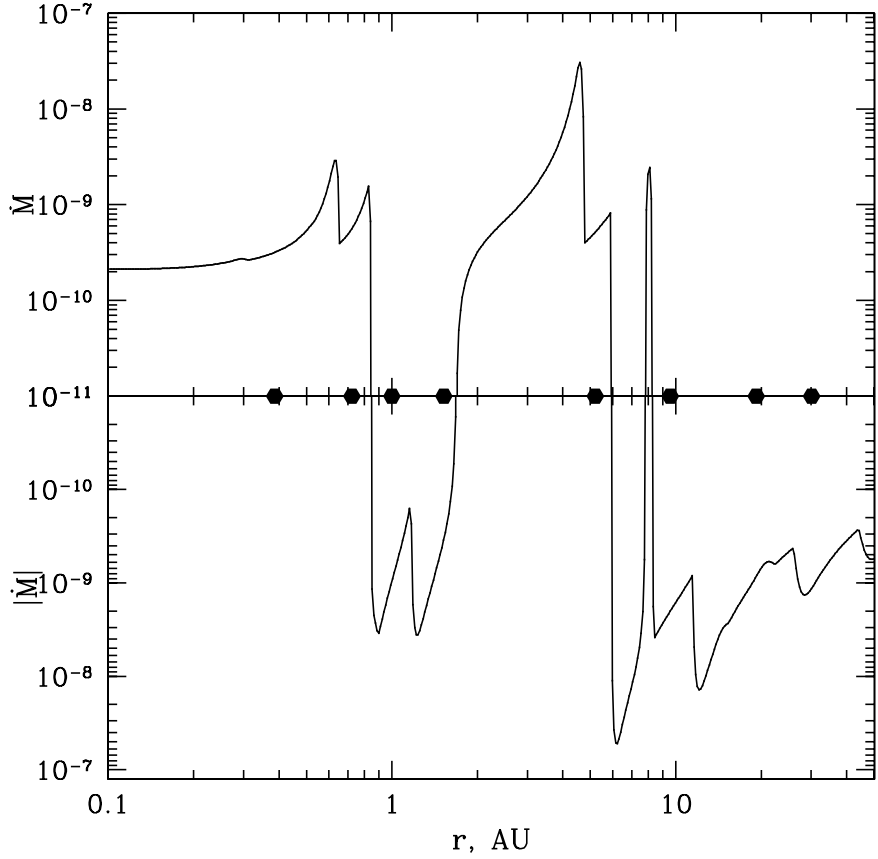}
\caption{Dependence of the planet-induced accretion rate $\dot M$ 
(in $M_\odot/$yr) upon the distance $r$ (in AU) in the minimum mass Solar Nebula (MMSN). Torques produced
by $8$ major planets (only masses of the rocky cores are taken for giants) 
are included here, and calculation is done using 
usual MMSN parameters [see equation (\ref{hayashi})]. 
Positive values of $\dot M$ (inflow towards the center) 
are displayed in the upper panel,
while negative ones (meaning the outflow) are in the lower panel.
Dots denote the locations of the planets. One
can see that in some positions in the nebula $\dot M$ could reach 
$>10^{-8}~M_\odot/$yr leading there to a significant surface density evolution
on timescales $\sim 10^6$ yr.
\label{Fig4}}
\end{figure}

In Fig. \ref{Fig4} we plot the accretion rate in the disk due
to the planetary torques from $8$ planets of the Solar System
(excluding Pluto)
using equation (\ref{tork}). It is assumed that
(Hayashi 1981)
\begin{equation}
\Sigma_0(r)=1700~ \mbox{g cm}^{-3}r_{AU}^{-3/2},~~~~ 
c_0(r)=1.2~ \mbox{km s}^{-1}r_{AU}^{-1/4}\label{hayashi}
\end{equation}
($r_{AU}$ is the distance from the center measured in AU),
and for the masses of the 
giant planets we take only the masses of their rocky cores: $15~M_\oplus$   
for Jupiter and Saturn and  $10~M_\oplus$ for Uranus and Neptune (otherwise 
they are likely to open gaps).

One can see that in some locations in the nebula
$\dot M\sim (1-5)\times 10^{-8}~M_\odot/$yr suggesting a significant
surface density evolution there. 
Indeed, the total disk mass inside Neptune's orbit
calculated using equation (\ref{hayashi}) is only $\approx 0.007~M_\odot$.
Averaging $\dot M$ over the bulk of the disk one obtains
$\langle\dot M\rangle\approx (2-4)\times 10^{-9}~ M_{\odot}$
(depending on whether one averages $\dot M$ or $|\dot M|$)
which implies the typical dispersal time of the MMSN by planetary
torques $\approx (2-3)\times 10^6$ yr, in rough agreement with 
observations (Hartmann et al. 1998). For more massive disks evolution
could be more rapid, because the mass contained in the planets is
increased (for a fixed disk metallicity): planets could be more massive,
meaning higher ratio $M_p/M_1$ 
[see equation (\ref{tork})], or simply be more numerous. This  
leads to stronger accretion so that the timescales 
$\la 10^6$ yr could be typical. 
Note however that at some point this increase in accretion rate is stopped
by the tendency of massive planets to open gaps in the disk. This could 
drastically reduce their influence on the nebular evolution.
Notice also that since $\dot M$ is very 
inhomogeneous radially, planetary-driven disk 
evolution must be highly time-dependent.

A very useful feature of our analysis presented in \S \ref{basder} 
is that it can be
applied to disks in which  the 
surface density and sound speed have arbitrary radial distributions. 
They should only vary on scales larger than the perturbation 
wavelength for our tight-winding approximation to be valid
(since in differentially rotating disks radial wavenumber
rapidly grows with distance from the planet this condition does not
pose serious restrictions). Thus one can not only calculate the instantaneous
accretion rate at each point in the disk but 
also study the self-consistent 
temporal evolution of the disk driven by the planetary torques. This is
very important, for example, for studying the gap formation around the planet
in gaseous disks and we are going to investigate this process in a future
study (Rafikov 2001).

\section{Conclusions.}\label{concl}

Tidal interaction of the gas disk with a planet embedded in it is
 important for the question of the
orbital evolution of the planet as well as for the fate of the  disk itself.
In this paper we presented a global description of the evolution of 
density waves in vertically isothermal disks where two-dimensional 
fluid equations provide proper approximation. 
The disk surface density and sound speed are allowed to 
vary independently with radius.

Our quantitative results are not very different from those obtained by
Goodman \& Rafikov (GR01): for low-mass planets surface density 
perturbations are weakly nonlinear and they shock after propagating several
local disk scale lengths $h_p$ from the planet. Subsequent damping of
the wave transfers its angular  momentum to the disk and is intrinsically 
asymmetric, which could be important for planet migration.
Disks evolve due to this angular momentum flux deposition 
and in the absence of other viscous mechanisms this could be
the only driver of their evolution. We have demonstrated that for the 
parameters similar to those of the Solar System the tidal 
perturbations alone could
produce spatially nonuniform  and time dependent evolution
with average accretion rates $\dot M\sim 10^{-9}-10^{-8}~M_\odot/$yr
(yielding typical timescale $\sim 10^6-10^7$ yr). 
In protoplanetary
systems with more favorable conditions disk evolution may be even stronger.

The prescription for the angular momentum deposition by planets 
in disks with arbitrary surface density and temperature profiles 
described here 
could easily be incorporated into other problems such as the gap formation 
around massive planet
or planet-driven global evolution of the surface density 
in the disk.

\section{Acknowledgements.} 

I am indebted to Jeremy Goodman and Scott Tremaine for illuminating
discussions and a lot of thoughtful comments on the manuscript. 
Financial support 
provided by the NASA Origins Program under grant NAG 5-8385 and NASA grant 
NAG 5-10456 is gratefully acknowledged.

\appendix

\section{Reduction of the equation for $R_-$ to Burgers equation.}\label{app1}

The equation for $R_-=u-2c/(\gamma-1)=2(c_0-2c)/(\gamma-1)$ reads
\begin{eqnarray}
\left[\partial_\phi+(u- c)\partial_\xi\right]R_-=
-\left[\frac{1}{\Sigma}\partial_\xi P-\frac{1}{\Sigma_0}\partial_\xi P_0
-c\partial_\xi\frac{2c}{\gamma-1}- cu\frac{\partial_\xi\Sigma}{\Sigma}
- u\partial_\xi\frac{2c}{\gamma-1}\right]\nonumber\\
+\frac{2\Omega}{\Delta \Omega} v +\frac{v^2}{\Delta \Omega r}+ 
\frac{c u}{\Delta \Omega r}+ \frac{c v}{\Delta \Omega r}
\partial_\phi\ln\Sigma
+ \frac{c}{\Delta \Omega r}\partial_\phi v.\label{Rminus}
\end{eqnarray}

We find it useful to introduce a new function
\begin{equation}
\psi\equiv\frac{\gamma+1}{\gamma-1}\frac{c-c_0}{c_0}\ll 1.\label{psidef}
\end{equation}
Using this definition and equation (\ref{sound_vel}) one can find that up 
to the second order in $\psi$
\begin{equation}
\frac{\Sigma-\Sigma_0}{\Sigma_0}=\frac{2}{\gamma+1}\psi+\frac{3-\gamma}
{(\gamma+1)^2}\psi^2. \label{dsig}
\end{equation}
Also from the conservation of the Riemann invariant $R_+$ we 
have that
\begin{equation}
u=2\frac{c_0-c}{\gamma-1}=-2\frac{c_0}{\gamma+1}\psi.\label{u}
\end{equation}  

Now, we want to relate $\partial_\xi P$ to the derivatives of 
$\Sigma-\Sigma_0$. 
When doing this we have to remember that 
relation (\ref{pol-law}) holds true only in the reference frame {\it comoving}
 with the fluid. Thus, Eulerian increment of any quantity $\Delta_E$
has to be related to its Lagrangian counterpart $\Delta_L$ by
$\Delta_E=\Delta_L-{\bf d}\nabla$, where ${\bf d}$ is 
a Lagrangian displacement.
Obviously, $\nabla\cdot{\bf d}=-
(\Sigma-\Sigma_0)/\Sigma_0$, and since the disk is
assumed to be thin and the 
spiral pattern is tightly wound, $\partial_\xi{\bf d}
\approx\nabla\cdot{\bf d}=-(\Sigma-\Sigma_0)/\Sigma_0$.
 Expanding Lagrangian increment of $P$
up to the second order in $\Sigma-\Sigma_0$ we obtain that
\begin{equation}
\partial_\xi P=\partial_\xi P_0+\partial_\xi\left[c^2(\Sigma-\Sigma_0)\right]
+\frac{\gamma-1}{2}\frac{c_0^2}{\Sigma_0}\partial_\xi
\left(\Sigma-\Sigma_0\right)^2
+P\frac{\Sigma-\Sigma_0}{\Sigma_0}\partial_\xi\ln\frac{P_0}{\Sigma_0^\gamma}.
\label{pderiv}
\end{equation}
For a polytropic equation of state with a fixed polytropic constant
(entropy in the disk is independent of $r$) the last
term in (\ref{pderiv}) is absent and the derivative of the pressure 
could be taken without worrying about the Lagrangian displacement. 

Since the tidal perturbation is assumed to be weak and tightly wrapped,
the most important 
nonlinear terms 
responsible for the wake evolution  are proportional to 
$\psi\partial_\xi\psi$ while terms like $\psi^2$ can be disregarded.
Substituting (\ref{dsig}), (\ref{u}), and (\ref{pderiv}) into (\ref{Rminus})
we get after lengthy but straightforward calculation that
\begin{eqnarray}
\partial_\phi\psi-c_0(1+\psi)\partial_\xi \psi=-\frac{\gamma+1}{4c_0}
\left[2\frac{\Omega}{\Delta \Omega}v+\frac{v^2}{\Delta \Omega r}+
\frac{c u}{\Delta \Omega r}+ \frac{c v}{\Delta \Omega r}
\partial_\phi\ln\Sigma \right.\nonumber \\
\left.+\frac{c}{\Delta \Omega r}\partial_\phi v
-\frac{6c_0^2}{\gamma+1}\psi\partial_\xi\ln c_0
-\frac{2c_0^2}{\gamma+1}\psi\partial_\xi\ln\Sigma_0\right].
\label{intres}
\end{eqnarray}
We needed to expand $\Sigma-\Sigma_0$ and $\partial_\xi P$ up to the second 
order in perturbation to make sure that all the terms 
proportional to $\partial_\xi \psi$ and $\psi\partial_\xi\psi$ cancel out
in the r.h.s 
of (\ref{Rminus}).

To an adequate approximation we can express
from (\ref{linphi1})  $\partial_\phi v$ as
\begin{equation}
\partial_\phi v\approx -\frac{2 B}{\Delta \Omega(r)} u.\label{parv}
\end{equation}
  Here we neglected
the nonlinear terms $u\partial_\phi v$ and $u v /(\Delta \Omega r)$; 
the term with 
$\partial_\phi \Sigma$ only slightly changes the propagation velocity of the 
wake and thus is disregarded too. 

In equation (\ref{linphi1}) 
the terms which are in phase with $u$ or $\psi$ are important for the 
amplitude of the perturbation evolution while those out-of-phase
only slightly affect the characteristic velocity but not the wave 
amplitude (as we noticed before). 
Derivative with respect to $\phi$ changes phase by $\pi/2$;
 thus, terms with $\partial_\phi$ in equation (\ref{linphi1}) can be 
considered separately from others. We can integrate them over $\phi$ to obtain
\begin{equation}
v\approx -2\frac{c_0^2}{\Delta \Omega r}\frac{1}{\gamma+1}\psi.
\label{v}
\end{equation}
This result, combined with equation (\ref{u}), confirms that $v\ll u$ (see \S
\ref{basder}).

Using relations (\ref{u}), (\ref{parv}), and (\ref{v})
we finally get that
\begin{eqnarray}
\partial_\phi\psi-c_0\left(1+\psi\right)
\partial_\xi\psi=\psi\frac{c_0}{\Delta \Omega r}
\left[\frac{1}{2} + 
\frac{1}{2}\frac{\partial\ln\Sigma_0}{\partial \ln r} 
+ \frac{3}{2}\frac{\partial\ln c_0}{\partial \ln r} -
\frac{A(r)}{\Delta \Omega}\right]+O(\psi^2).\label{r-minus}
\end{eqnarray}
The 
local approximation of GR01 could be retrieved now by assuming $\Sigma_0=const$
and $c_0=const$ and expanding $\Delta \Omega$ to first order in terms of 
$r-r_p$.

All effects of the nonlinear wake evolution 
are embodied  in the term $\psi\partial_\xi\psi$ in the l.h.s. of
equation (\ref{r-minus}).
We now reduce this equation to the conventional Burger's equation
following the approach outlined in GR01. First, we make a 
change of independent variables from $\phi, \xi$ to 
$\phi^\prime, \eta_1$ given by the relations
\begin{eqnarray}
\int\frac{d\xi}{c(\xi)}\to \eta_1-\phi^\prime,~~~~~~\phi\to\phi^\prime
\label{varch}
\end{eqnarray}
Here $\phi^\prime$ has the meaning of the azimuthal 
distance along the wake while $\eta_1$ represents the displacement
from the wake center in the $\phi$-direction.
Since the wake is narrow, $\eta\ll\phi$ and we can consider $c_0$ to be a function of $\phi^\prime$ only: $c_0(\xi)=c_0(\phi)$.
This transforms equation (\ref{r-minus}) into 
(we drop the $\prime$ from $\phi^\prime$)
\begin{equation}
\partial_\phi\psi-\psi\partial_{\eta_1}\psi=
-\psi \frac{1}{g(\phi)}\partial_\phi g(\phi),\label{inteq}
\end{equation}
where function $g(\phi)$ is defined by
\begin{equation}
\frac{1}{g(\phi)}\partial_\phi g(\phi)=-\frac{c_0(\phi)}{\Delta \Omega(\phi) 
r(\phi)}
\left[\frac{1}{2} + 
\frac{1}{2}\frac{\partial\ln\Sigma_0}{\partial \ln r} 
+ \frac{3}{2}\frac{\partial\ln c_0}{\partial \ln r} -
\frac{A(\phi)}{\Delta \Omega(\phi)}\right].\label{gdef}
\end{equation}

Introducing new function $\chi(\phi)=g(\phi)\psi$ we rewrite the equation 
(\ref{inteq}) as
\begin{equation}
\partial_\phi \chi=\chi\frac{1}{g(\phi)}\partial_{\eta_1} 
\chi.\label{penult}
\end{equation}

Changing from $\phi$ to a new variable $t_1$ given by
\begin{equation}
t_1=\int_0^\phi\frac{d\phi^\prime}{g(\phi^\prime)}\label{t}
\end{equation}
we arrive at the Burger's equation
\begin{eqnarray}
\partial_{t_1}\chi-\chi\partial_{\eta_1}\chi=0.\nonumber
\end{eqnarray}

Both $t_1$ and $\eta_1$ are dimensionless and we find it useful to 
rescale them to $t, \eta$ in the following way:
\begin{equation}
\eta=\frac{r_p}{l_p}\eta_1,~~~t=\frac{r_p}{l_p} t_1,\label{renorm}
\end{equation}
where $l_p=c_p/|2A(r_p)|$ is a Mach-1 length 
at the position of the
planet (in the limit $r\to r_p$ this rescaling makes our $\eta$
identical with $\eta$ used in GR01). Thus, we obtain equation (\ref{burg}).

Finally, we calculate the behavior of function $g$.
From definitions (\ref{xi}) and (\ref{varch}) one finds that 
$d\phi=-(\Delta\Omega/c_0)dr$. Using this and the definition of
Oort's parameter $A$ one can easily integrate equation (\ref{gdef}) to find
\begin{equation}
g=C_1\left(\frac{r\Sigma_0 c_0^3}{|\Delta\Omega|}\right)^{1/2},
~~~C_1=const.\label{g_fin}
\end{equation}
The arbitrary constant $C_1$ should be chosen in such a way
 that in the limit $r\to r_p$ our definitions of $\chi$ and $t$
reduce to the analogous expressions of GR01 [given by equation (\ref{local})]. 
One achieves this by taking
\begin{equation}
C_1=\frac{2^{1/4}}{r_p c_p \Sigma_p^{1/2}}
\end{equation}
[see equations (\ref{g})-(\ref{local})].

\end{document}